\documentclass{icrc2009}

\usepackage{graphicx}   
\usepackage{caption}    
\usepackage[font=footnotesize]{subfig} 
\usepackage{fixltx2e}
\usepackage{url}

\newcommand{\shorttitle}[1]%
{\markboth{Proceedings of the 31\MakeLowercase{$^{st}$} ICRC, {\L}\'{o}d\'{z} 2009}{#1} }
\newcommand{\etal}{\MakeLowercase{\textit{et al. }}} 

\begin{document}
\title{Experimental search of bursts of gamma rays from primordial black holes
using different evaporation models}

\author{\IEEEauthorblockN{
E.V. Bugaev\IEEEauthorrefmark{1}, V.B. Petkov\IEEEauthorrefmark{1}\IEEEauthorrefmark{2},
A.N. Gaponenko\IEEEauthorrefmark{1}, P.A. Klimai\IEEEauthorrefmark{1},
M.V. Andreev\IEEEauthorrefmark{1}\IEEEauthorrefmark{2}\IEEEauthorrefmark{3},
A.B. Chernyaev\IEEEauthorrefmark{1}, \\ I.M. Dzaparova\IEEEauthorrefmark{1}\IEEEauthorrefmark{2}, D.D. Dzhappuev\IEEEauthorrefmark{1},
Zh.Sh. Guliev\IEEEauthorrefmark{1}, N.S. Khaerdinov\IEEEauthorrefmark{1}, N.F. Klimenko\IEEEauthorrefmark{1}, \\
A.U. Kudzhaev\IEEEauthorrefmark{1}, A.V. Sergeev\IEEEauthorrefmark{1}\IEEEauthorrefmark{2}\IEEEauthorrefmark{3},
V.I. Volchenko\IEEEauthorrefmark{1}, G.V. Volchenko\IEEEauthorrefmark{1} and A.F. Yanin\IEEEauthorrefmark{1} }
                            \\
\IEEEauthorblockA{ \IEEEauthorrefmark{1}
 Institute for Nuclear Research, Russian Academy of Sciences, \\
 60th October Anniversary Prospect 7a, 117312 Moscow, Russia \\
 \IEEEauthorrefmark{2}
 Terskol Branch of the Institute of Astronomy of the Russian Academy of Sciences \\
 \IEEEauthorrefmark{3}
 International Center for Astronomical,  Medical and Ecological Research, \\
 National Academy of Sciences of Ukraine
 }
}

\shorttitle{Bugaev \etal Experimental search of bursts...}
\maketitle

\begin{abstract}

Experimental data of arrays "Andyrchy" and "Carpet-2" of Baksan Neutrino Observatory
(Institute for Nuclear Research), obtained in the regime of a detection of the single
cosmic-ray component, are used for a search of the bursts of cosmic gamma rays from
evaporating primordial black holes. Different theoretical models of the evaporation process
are used for the analysis. Distributions of the counting rate fluctuations on both
arrays agree with the expectations from the cosmic ray background. The new constraints
on the concentration of evaporating primordial black holes in the local region of Galaxy
are obtained. The comparison of the results of different experiments is given.

\end{abstract}


\section{Introduction}

It has been argued recently \cite{MacGibbon:2007yq} that the photon flux calculation from an evaporating
black hole (BH) given in \cite{MacGibbon:1990zk} is the most reliable, and the photospheric
and chromospheric effects considered in \cite{Heckler}, \cite{Kapusta} are negligibly small.
It is clear, however (and we try to show this in this section), that scenarios of BH evolution,
in which interactions between emitted particles (and even some kind of the thermal atmosphere around
the BH) exist, cannot, in general, be discarded.

In Secs. \ref{sec-exp} and \ref{sec-res} we study the detection of BH evaporation using three models
(considering them on an equal footing). We do not insist on the validity of concrete chromospheric models;
our aim is to demonstrate the sensitivity of the experimental method used in the present paper
(which is based on the registration of evaporated photons with energies $\sim 10$ GeV) to a form of the BH
photon spectrum, having in mind that its right shape may be different from the "canonical" one.

The famous Hawking's result \cite{Hawking:1974sw} according to which a BH will emit black body radiation,
precisely corresponding to the temperature $T_H= \frac{T_R}{4G m_{BH}}$ (where $T_R$ is the
dimensionless Rindler temperature, $T_R=\frac{1}{2\pi}$) had been obtained in a semiclassical
approximation. It had been assumed, in particular, that back-reaction effects are effectively
small and do not influence the classical collapse geometry. It is well known that Hawking's
derivation of the evaporation spectrum is based on information loss phenomenon \cite{Hawking:1976ra} and,
correspondingly, on the possibility of an nonunitary evolution of the BH. This is a direct consequence
of the strict locality of quantum field theory which leads to an independence of the external
and internal Hilbert states (i.e., states located outside and inside horizon, respectively)
and, as a result, to an appearing of the density matrix for the BH decay rather than a pure state.

It had been argued in many works, however (see, e.g.,
\cite{'tHooft:1996tq, Giddings:2004ud, Giddings:2006sj}), that the semiclassical
approximation used in \cite{Hawking:1974sw} is too crude: generically, one cannot regard outgoing particle
as a Hawking particle that was not affected by the ingoing matter. Moreover,
the gravitational interactions of the Hawking radiation (when it already separates from
the horizon) with infalling particles can be strong enough for a deforming the metric and for a
breakdown of locality. It is known that the dynamics of strong gravitational physics is
inherently nonlocal, and it is natural to assume that just this nonlocality may be the
resolution of the above-mentioned paradox of information loss \cite{Giddings:2006sj}.

An approximate (effective) accounting of the backreaction effects is realized in the approach in which
the boundary condition for all fields on a surface a few Planck distances away from horizon is postulated.
The examples are the "brick wall" model \cite{'tHooft:1984re}, the "stretched horizon" model
\cite{Susskind:1993if} and the "bounce" model \cite{Stephens:1993an}. These models are based on
an assumption that strong gravitational interactions among the field quanta effectively form this barrier
between the horizon and the wall.

It is argued in these works that the process of formation and evaporation of a BH, as viewed by
a distant observer, can be described entirely within the context of standard quantum theory,
with unitary $S$-matrix and pure quantum states.

An idea of the stretched horizon (SH) was proposed many years ago \cite{Thorne:1986iy} but at that times it
was considered as a useful mathematical construction rather than a physical object equipped with some
microphysical degrees of freedom. One should stress, however, that the approach using a concept
of the SH is appropriate only for an external observer, not for a free falling one. The main point
of the approach is that the SH can, through its degrees of freedom, absorb, thermalize and emit any
quantum mechanical information falling into the BH, without any information loss \cite{Susskind:1993if}.

The local proper temperature at the SH, $T_s$, is related to the temperature measured by distant observers
(which is, by definition, the Hawking temperature $T_H$), by the connection
\begin{equation}
T_s = \frac{dt}{d\tau} T_H, \label{Ts}
\end{equation}
where $d\tau/dt$ is the time dilation factor, connecting the time intervals at the
SH with the intervals of coordinate time $t$, which is given by
\begin{equation}
\frac{d\tau}{dt} = \frac{\rho_h}{4 M G} \sim \frac{\rho_h}{r_h}. \label{dtaudt}
\end{equation}
Here, $\rho_h$ is the distance of SH from the horizon, $r_h$ is the gravitational radius.
In the "standard" case, when $\rho_h \sim G M_P \equiv l_P$ ($M_P$ is the Planck mass), one
has
\begin{equation}
\frac{d\tau}{dt} \sim \frac{M_P}{M},
\end{equation}
and, if $T_s \sim M_P$ (independently of the size or mass of the BH), one obtains
\begin{equation}
T_H \sim \frac{M_P^2}{M}.
\end{equation}

It is very important that the SH is assumed to be in a thermal equilibrium with the surrounding material during
most of the evaporation. A distant observer would estimate the number of particles emitted per unit
time which is proportional to a product of the BH area and the time dilation factor \cite{Susskind:1993if},
\begin{equation}
\frac{dN}{dt}\sim M^2 \frac{d\tau}{dt} \sim M \label{e2}
\end{equation}
(if all these particles go to infinity). On the other hand, the number per unit time of particles
that actually emerged to infinity is
\begin{equation}
\frac{dN}{dt}\sim \frac{L}{E_{\rm typ}} \sim \frac{1}{M} \label{e3}
\end{equation}
($E_{\rm typ}$ is the typical energy of the emitted particles, $E_{\rm typ}\sim T_H\sim M_P^2/M$,
$L$ is the BH luminosity, $L\sim 1/M^2$).
It follows from (\ref{e2}) and (\ref{e3}) that most of the particles emitted from
the SH do not go to infinity. This gives rise to a thermal atmosphere above
the SH (due to repeated interactions of emitted particles with the SH and with each other).

One should stress that, in these models, BH decays as a pure quantum mechanical state, so,
a spectrum of the emitted radiation does not need to have a precise thermal form
predicted in \cite{Hawking:1974sw}, at least at final stages of the evaporation, when
BH mass is small.

In general case, the temperature of the SH, $T_s$, as well as the distance $\rho_h$,
depend on microphysics. Correspondingly, the maximum value of the Hawking temperature $T_H$,
and a form of the spectrum of evaporated particles are model dependent. Evidently, this model
dependence is especially large at final stages of the evaporation when a radius of the
BH, $r_h$, is comparable with $\rho_h$. In this case the centrifugal barrier becomes
inefficient and the thermal atmosphere around the SH gradually becomes open for a space
surrounding the BH. The experimental signature may be similar with those predicted
in chromospheric models if the maximum Hawking temperature $T_H$ is much smaller than $M_P$,
as is predicted in some scenarios. For example, such a situation is possible in string models.
In these models, the SH is placed at a distance $\sim l_s$ from the event horizon ($l_s$ is the string
scale). The local proper temperature at the SH is equal to \cite{Susskind:1993ws} $T_s=1/2\pi l_s$.
We suppose, that the string coupling, $g^2=G l_s^2$, is extremely small, so that
the Planck and string scales are well separated. When, in a course of the evaporation,
the Hawking temperature $T_H$ approaches $T_s$ (which is the Hagedorn temperature
of string theory), the radius of the BH becomes equal to $l_s$. In this point
the BH can transform into a higher-entropy string state. The possibility of
such transformation was discussed in many works (see \cite{Damour:1999aw}
and references therein).

It is essential also that the ideas of the string-BH correspondence \cite{Horowitz:1996nw}
and of identifying the states of a BH with the highly excited states of a fundamental
string (\cite{Susskind:1993ws, Halyo:1996xe}) help us to understand the physical nature of the BH entropy
(by other words, the nature of the "internal states of a BH", or, "the degrees of
freedom of the stretched horizon").

It is important to note that strings are ideally suited for interactions with the SH due
to a spread in the transverse space (during approaching the SH) and due to Regge behavior
of string scattering amplitudes and their growing cross sections \cite{Susskind:1993aa}.

The most important conclusion from this scenario is that the resulting, weakly
coupled, string decays at the Hagedorn temperature, and the momenta of the emitted
particles never exceed $T_s$ (which can be as small as $0.1$ TeV).

\section{The experiment} \label{sec-exp}

Up to now, the search for the bursts of high-energy gamma rays which are generated at the final stages of PBH evaporation
was carried out in ground-based experiments detecting the EASs on several shower arrays \cite{Connaughton:1998ac, Alexandreas:1993zx,
Amenomori, Funk, Petkov:2008rz} and on the Whipple Cherenkov telescope \cite{Linton:2006yu}.
Because of high energy detection threshold, the interpretation of the results of these
experiments can be performed only within the framework of the evaporation model without a chromosphere \cite{MacGibbon:1990zk}
(hereafter, MW90). The duration of the high-energy burst predicted by chromospheric evaporation models is too short, much
shorter than the detection dead time in these experiments (the duration of burst $t_b$ is, by definition, the time interval during which
99\% of photons that can be detected by the array evaporate).

\begin{figure}
\centering
\includegraphics[width=0.96 \columnwidth ]{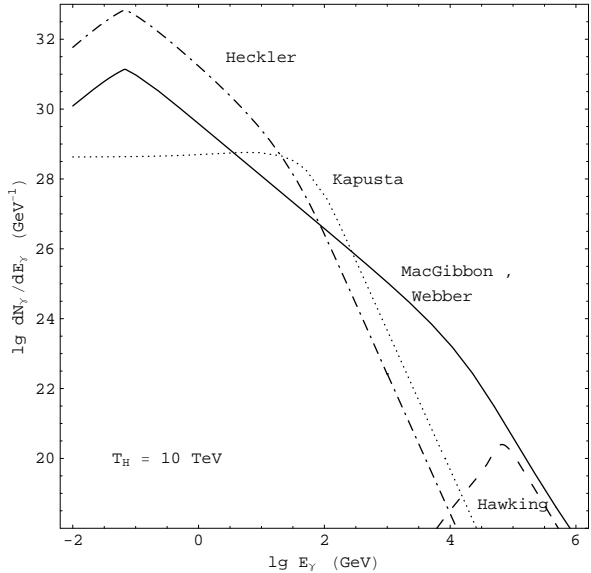}
\caption{Time-integrated photon spectra from a BH with initial Hawking temperature $T_H = 10$ TeV.
Such a BH has about 0.5 s until the full evaporation.}
\label{fig1}
\end{figure}

For the direct search of PBH gamma ray bursts within the framework of the chromospheric models of BH evaporation,
another technique has to be applied, which is sensitive to photons with rather small energies, namely of order of few GeV.
Shower particles generated in atmosphere by photons with energy
$\sim 10 - 100$ GeV are strongly absorbed before reaching the detector
level, so, the average number of signals in the detector
module is smaller than 1. Therefore, in this energy range PBH bursts
can be sought for by operating with the modules in single
particle mode, that is, by measuring the single particle counting rate
of the individual modules. The primary arrival directions of
photons are not measured, and bursts can be detected only as spikes
(short-time increases) of the cosmic ray counting rate.
The effective energy of the primary gamma-ray photons detected by
this method depends mainly on the altitude of the array above sea
level. This technique was employed earlier to seek for cosmic gamma-ray bursts with energies exceeding several GeV
\cite{EASTOP, Vernetto, Alexeenko, Petkov2003}. First constraints on the number density of evaporating PBHs within
chomospheric framework were obtained in work \cite{Petkov:2008rv} for the evaporation models by Heckler \cite{Heckler}
(hereafter, H97) and Daghigh and Kapusta \cite{Kapusta} (hereafter, DK02). It should be noted that for the PBH search
within the framework of model MW90, the same technique still can be applied. The comparison of three evaporation
models considered can be found in \cite{Bugaev:2007py}, and time-integrated photon spectra for particular value of Hawking
temperature are shown in Fig. \ref{fig1}.

The present experiment was performed on Andyrchy and Carpet-2 arrays of Baksan neutrino observatory of INR RAS.
Both arrays are situated close to each other (horizontal distance between them is about 1 km), but their altitudes
above sea level are different: Andyrchy - 2060 m (atmospheric depth - 800 g cm$^{-2}$), Carpet - 1700 m
(atmospheric depth - 840 g cm$^{-2}$).

Andyrchy consists of 37 scintillation detectors. The area of each detector is 1 m$^2$. To detect a
single cosmic-ray component, the total counting rate of all of the detectors is measured each second. The search
for gamma-ray bursts using this technique is carried out
against the high cosmic-ray background ($\bar \omega = 11440 \;{\rm s}^{-1}$),
which requires the highly stable and reliable performance of all equipment. The monitoring is realized
through simultaneous measurements (with 1-s acquisition rate) of the counting rates in the four parts of the array
comprising 10, 9, 9, and 9 detectors, respectively. An use of such information allows to exclude 1-second intervals
with unreasonably large deviations of counting rate between the parts of the array which could come from
faultiness of individual detectors or registration channels, faultiness of counting rate measurement channels
of separate parts of the array or non-synchronous impulsive electromagnetic disturbance in detectors
or signal cables.
A detailed description of the array and its operating parameters is given in \cite{Petkov:2006zz}.

Carpet-2 array \cite{Alekseev:1976ac, Dzhappuev} consists of the actual Carpet facility
(400 liquid scintillation detectors covering area of 196 m$^2$),
six remote points (RP; 108 detectors of the same kind with total area 54 m$^2$), and the muon detector
(175 detectors of the type used in Andyrchy array placed in the underground tunnel).

The detection probabilities $P(E_\gamma, \theta)$ of the secondary particles
produced by the primary photons with energy $E_\gamma$ falling on the arrays at zenith angle
$\theta$ were determined by simulating electromagnetic cascades in the atmosphere and the detectors.
The effective energy of gamma rays registered by Andyrchy and RP is 8 GeV, for Carpet and muon
detector this energy is, correspondingly, 200 GeV and 2 TeV. Because of this, only data from
Andyrchy and RP was actually used in this work.

The total number of gamma rays which can be detected by the array is given by
\begin{equation}
N_\gamma(\theta, t_l) = \int\limits_0^\infty dE_\gamma P(E_\gamma, \theta) dN_\gamma/dE_\gamma.
\end{equation}
It depends on the time $t_l$ left until the end of PBH evaporation. Here, $dN_\gamma/dE_\gamma$
is the spectrum of photons emitted by the PBH during the same time interval $t_l$.
Let a PBH be located at distance $r$ from the array with area $S$ and be seen from it at zenith
angle $\theta$. The mean number of gamma-rays detected by the array is then
\begin{equation}
\bar n(\theta, r) = \frac{N_\gamma(\theta, t_l)S(\theta)}{4\pi r^2}\; ,
\label{eq2}
\end{equation}
which will give an excess over the average counting rate $\bar \omega$ in
units of root mean square deviation:
\begin{equation}
f(\theta) = \frac{\bar n(\theta, r)} {\sqrt{\bar \omega} } = \frac{N_\gamma(\theta, t_l)S(\theta)}{4\pi r^2 \sqrt{\bar \omega}} \;.
\end{equation}

\begin{figure}[!t]
\centering
\includegraphics[width=0.99 \columnwidth ]{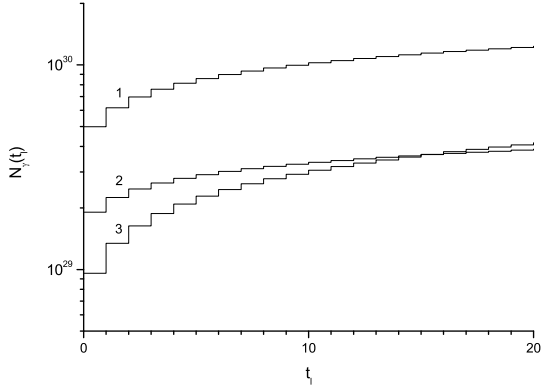}
\caption{Full number of gamma particles that can be detected by the Andyrchy array for $\theta=0^\circ$,
as a function of time until the end of PBH evaporation, for different evaporation models: 1 - DK02, 2 - H97,
3 - MW90. }
\label{fig2}
\end{figure}

The time interval $\Delta t$ which should be used to search for bursts of gamma rays from evaporating PBHs in
the particular experiment depends both on the burst duration and on its time profile. Burst duration
for Andyrchy and RP ($E_{th} \sim 8$ GeV) exceeds $10^4$ s for all considered evaporation models.
It should be noted that the search of long bursts (tens of seconds and more) in ground-based experiment,
especially when working in single-particle mode, is difficult because of short-time variations of cosmic
ray intensity \cite{Vernetto, Petkov2003}. Fig. \ref{fig2} shows, for considered models, the dependencies of
total number of gamma rays that can be detected by Andyrchy array on the time $t_l$. It is seen that
even for a rather strong increase of $t_l$ (from 1 s to 20 s) the number of gamma particles increases
only by a factor of 2 to 4 (depending on the model). At the same time, the number of background events
increases proportional to the time interval. The consequence of this consideration is that in our experiment,
the search is most effective for 1-second time intervals.

\begin{figure}[!t]
\centering
\includegraphics[width=0.99 \columnwidth ]{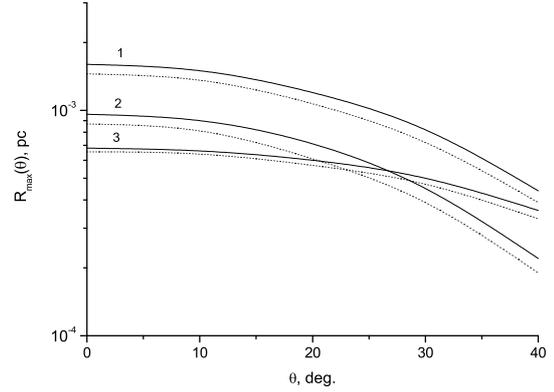}
\caption{ Maximum distance from which the evaporating PBH can be detected.
Solid lines are for Andyrchy (number of particles in this case is $n=640$),
dotted lines are for RP ($n=918$); models used are: 1 - DK02, 2 - H97,
3 - MW90. }
\label{fig3}
\end{figure}

Deviations in the array counting rate lasting $\Delta t \le 1$ s are sought using the parameter $F_i$ that is equal to the deviation
(measured in units of the Poisson sigma) of the number of counts $k_i$ during the $i$-th second of a 15-min
interval from the average number of counts during this interval: $F_i=(k_i-\bar k)/\sqrt{\bar k}$. Since variations in the
cosmic-ray intensity over a time of 15 min are negligible in the first approximation and the average counting rate
is fairly high, one can expect that the parameter $F_i$ has a Gaussian distribution with the zero mean value $V = 0$
and unit standard deviation $\sigma = 1.0$.

The actual experimental data mostly confirms this assumption. For the RP data, no excesses were found larger than 7
standard deviations. For Andyrchy, the only event with a large ($7.9\sigma$) deviation was detected on April 17, 2002 at 17:31:29 UT
, all other data is well fitted by a Gaussian distribution with maximum deviations of $6\sigma$.
If we assume that the $7.9\sigma$ Andyrchy event
is caused by evaporating PBH, then, taking into account that RP is examining the same region in the sky, the excess in
RP counting rate should be in the region  $(6.6-7.7)\sigma$ for chromospheric models, or in the region $(8.2-8.5)\sigma$
for MW90 model. However, for this 15-min interval, no excess larger than $3.2\sigma$ was detected by RP, so this single
Andyrchy event cannot be explained by the evaporating PBH.

One can estimate the maximum distance from which a PBH can be seen using the formula [see Fig. \ref{fig3}]
\begin{equation}
R_{\rm max} = \sqrt{\frac{N_\gamma(\theta, t_l=1 \;{\rm s}) S(\theta)}{4\pi n}} \; .
\end{equation}

\section{Results and conclusions} \label{sec-res}

Generally, the effective volume surveyed by the array is calculated using the formula
\begin{equation}
V_{\rm eff}=\int d \Omega {\int \limits_{0}^{\infty} dr
r^2 F(n, \bar n(\theta, r))}.
\end{equation}
Here, $F(n,\bar n(\theta, r))$ is the Poisson probability to register $n$ or more events
having the average $\bar n(\theta, r)$ determined by Eq. (\ref{eq2}). Taking into account that no excesses above $6\sigma$ were
detected on Andyrchy, we put $n=6\sqrt{\bar\omega}=640$, and the effective volume
for this array is, depending on the model: $1.2 \times 10^{-10}$ pc$^{3}$,
$5.6 \times 10^{-10}$ pc$^{3}$, $8.1 \times 10^{-11}$ pc$^{3}$ for H97, DK02 and MW90, correspondingly.
For RP, no excesses above $7\sigma$ are discovered (in this case $\bar\omega=17200$, so, $n=7\sqrt{\bar\omega}=918$),
and the effective volume is: $7.6 \times 10^{-11}$ pc$^{3}$, $4.1 \times 10^{-10}$ pc$^{3}$, $7.1 \times 10^{-11}$ pc$^{3}$
for H97, DK02 and MW90, correspondingly.

The number of bursts detected over the total observation time $T$ can be represented as
\begin{equation}
N=\rho_{\rm pbh} T V_{\rm eff} \; ,
\end{equation}
where $\rho_{\rm pbh}$ is the number density of evaporating PBHs.
Assuming that evaporating PBHs are distributed uniformly in the local region of the Galaxy and taking into account that
both arrays survey the same sky region at different time, one can calculate the upper limit
$\rho_{\rm lim}$ on the number density of evaporating PBHs at the 99\% confidence level using the formula
\begin{equation}
\rho_{\rm lim} = 4.6 / (V_{A}T_{A} + V_{RP}T_{RP}),
\end{equation}
with full observation time $T_{A} = 6.27$ yr for Andyrchy and $T_{RP}=2.34$ yr for RP.
Substituting the effective volumes with values calculated above, we finally obtain
for $\rho_{\rm lim}$: $10^{9}$ pc$^{-3}$ yr$^{-1}$, $5 \times 10^{9}$ pc$^{-3}$ yr$^{-1}$,
$6.8 \times 10^{9}$ pc$^{-3}$ yr$^{-1}$ for evaporation models DK02, H97 and MW90, respectively.

One can note, that in a case of the non-chromospheric model MW90, the limit obtained in
this work is worse than the limit from the Andyrchy array obtained using method of EAS detection \cite{Petkov:2008rz}.
However because the effective energies of detected photons differ by several orders of magnitude, these limits
should be regarded as complementary. For the case of DK02 and H97 models, the limit has been
significantly improved compared to our previous work \cite{Petkov:2008rv}.

{\bf Acknowledgements.} This work was supported by the the Russian Foundation for Basic Research
(Grants No. 06-02-16135, 08-07-90400 and 09-02-90900). This work was also supported in part by the
"Neutrino Physics and Astrophysics" Program for Basic Research of the Presidium of the
Russian Academy of Sciences and by "State Program for Support of Leading Scientific Schools"
(Project No. NSh-321.2008.2).


\begin{thebibliography}{99}


\bibitem{MacGibbon:2007yq}
  J.~H.~MacGibbon, B.~J.~Carr and D.~N.~Page,
  Phys.\ Rev.\  D {\bf 78}, 064043 (2008)
  [arXiv:0709.2380 [astro-ph]].

\bibitem{MacGibbon:1990zk}
  J.~H.~MacGibbon and B.~R.~Webber,
  Phys.\ Rev.\  D {\bf 41}, 3052 (1990).

\vfill \pagebreak

\bibitem{Heckler}
A.\,F. Heckler, Phys.Rev. Lett. {\bf 78}, 3430 (1997).

\bibitem{Kapusta}
R.\,G. Daghigh and J.\,I. Kapusta, Phys.Rev. D {\bf 65}, 064028 (2002).

\bibitem{Hawking:1974sw}
  S.~W.~Hawking,
  Commun.\ Math.\ Phys.\  {\bf 43}, 199 (1975)
  [Erratum-ibid.\  {\bf 46}, 206 (1976)].

\bibitem{Hawking:1976ra}
  S.~W.~Hawking,
  Phys.\ Rev.\  D {\bf 14} (1976) 2460.


\bibitem{'tHooft:1996tq}
  G.~'t Hooft,
  Int.\ J.\ Mod.\ Phys.\  A {\bf 11}, 4623 (1996)
  [arXiv:gr-qc/9607022].

\bibitem{Giddings:2004ud}
  S.~B.~Giddings and M.~Lippert,
  Phys.\ Rev.\  D {\bf 69}, 124019 (2004)
  [arXiv:hep-th/0402073].

\bibitem{Giddings:2006sj}
  S.~B.~Giddings,
  Phys.\ Rev.\  D {\bf 74}, 106005 (2006)
  [arXiv:hep-th/0605196].

\bibitem{'tHooft:1984re}
  G.~'t Hooft,
  Nucl.\ Phys.\  B {\bf 256}, 727 (1985).


\bibitem{Susskind:1993if}
  L.~Susskind, L.~Thorlacius and J.~Uglum,
  Phys.\ Rev.\  D {\bf 48}, 3743 (1993)
  [arXiv:hep-th/9306069].

\bibitem{Stephens:1993an}
  C.~R.~Stephens, G.~'t Hooft and B.~F.~Whiting,
  Class.\ Quant.\ Grav.\  {\bf 11}, 621 (1994)
  [arXiv:gr-qc/9310006].

\bibitem{Thorne:1986iy}
  K.~S.~Thorne, R.~H.~Price and D.~A.~Macdonald,
  ``BLACK HOLES: THE MEMBRANE PARADIGM,''
{\it  NEW HAVEN, USA: YALE UNIV. PR. (1986) 367p}.

\bibitem{Susskind:1993ws}
  L.~Susskind,
  arXiv:hep-th/9309145.

\bibitem{Damour:1999aw}
  T.~Damour and G.~Veneziano,
  Nucl.\ Phys.\  B {\bf 568}, 93 (2000)
  [arXiv:hep-th/9907030].

\bibitem{Horowitz:1996nw}
  G.~T.~Horowitz and J.~Polchinski,
  Phys.\ Rev.\  D {\bf 55}, 6189 (1997)
  [arXiv:hep-th/9612146].

\bibitem{Halyo:1996xe}
  E.~Halyo, B.~Kol, A.~Rajaraman and L.~Susskind,
  Phys.\ Lett.\  B {\bf 401}, 15 (1997)
  [arXiv:hep-th/9609075].


\bibitem{Susskind:1993aa}
  L.~Susskind,
  Phys.\ Rev.\  D {\bf 49}, 6606 (1994)
  [arXiv:hep-th/9308139].

\bibitem{Connaughton:1998ac}
  V.~Connaughton {\it et al.},
  Astropart.\ Phys.\  {\bf 8}, 179 (1998).


\bibitem{Alexandreas:1993zx}
  D.~E.~Alexandreas {\it et al.},
  Phys.\ Rev.\ Lett.\  {\bf 71}, 2524 (1993).

\bibitem{Amenomori} M.Amenomori \etal, Proc. 24th International Cosmic Ray Conference, Rome (Italy), v.2, 112 (1995).

\bibitem{Funk} B.Funk \etal, Proc. 24th International Cosmic Ray Conference, Rome (Italy), v.2, 104 (1995).

\bibitem{Petkov:2008rz}
  V.~B.~Petkov {\it et al.},
  Astron.\ Lett.\  {\bf 34}, 509 (2008)
  [Pisma Astron.\ Zh.\  {\bf 34}, 563 (2008)]
  [arXiv:0808.3093 [astro-ph]].

\bibitem{Linton:2006yu}
  E.~T.~Linton {\it et al.},
  JCAP {\bf 0601}, 013 (2006).

\bibitem{EASTOP}
M. Aglietta \etal, 
Astrophys. J. {\bf 469}, 305-310 (1996).

\bibitem{Vernetto}
S.~Vernetto,
  Astropart.\ Phys.\  {\bf 13}, 75 (2000)
  [arXiv:astro-ph/9904324].

\bibitem{Alexeenko}
V. V. Alexeenko \etal, Nucl. Phys. (Proc. Suppl.) {\bf B 110}, 472 (2002).

\bibitem{Petkov2003}
V.~B.~Petkov \etal, 
Kinematics and physics of celestial bodies {\bf 3}, 234 (2003).


\bibitem{Petkov:2008rv}
  V.~B.~Petkov, E.~V.~Bugaev, P.~A.~Klimai and D.~V.~Smirnov,
  JETP Lett.\  {\bf 87}, 1 (2008)
  [Pisma Zh.\ Eksp.\ Teor.\ Fiz.\  {\bf 87}, 3 (2008)]
  [arXiv:0803.2313 [astro-ph]].




\bibitem{Bugaev:2007py}
  E.~Bugaev, P.~Klimai and V.~Petkov,
  Proc. 30th ICRC, Mexico , v.3, 1123 (2007).
  arXiv:0706.3778 [astro-ph].

\bibitem{Petkov:2006zz}
  V.~B.~Petkov {\it et al.},
  Instrum.\ Exp.\ Tech.\  {\bf 49}, 785 (2006).

\bibitem{Alekseev:1976ac}
  E.~N.~Alekseev {\it et al.},
  Izv.\ Akad.\ Nauk Ser.\ Fiz.\  {\bf 40}, 994 (1976).

\bibitem{Dzhappuev}
 D.~D.~Dzhappuev \etal, Bull. Rus. Acad. Sci.: Physics {\bf 71} N4, 525 (2007).





\end{thebibliography}
\end{document}